\documentclass[universe,article,accept,pdftex,moreauthors]{Definitions/mdpi}
\usepackage{braket}
\usepackage{amsmath}
\usepackage{color}
\usepackage{xcolor}

\firstpage{1} 
\pubvolume{1}
\issuenum{1}
\articlenumber{0}
\pubyear{2025}
\copyrightyear{2025}
\externaleditor{Firstname Lastname} 
\datereceived{15 June 2025} 
\daterevised{25 July 2025} 
\dateaccepted{28 July 2025} 
\datepublished{ } 
\hreflink{https://doi.org/} 
\Title{Nuclear Matter and Finite Nuclei: Relativistic Thomas--Fermi 
	Approximation 
	Versus Relativistic Mean-Field Approach}

\TitleCitation{Nuclear Matter and Finite Nuclei: Relativistic Thomas--Fermi Approximation 
	Versus Relativistic Mean-Field Approach}

\Author{Shuying Li, Hong Shen *\orcidA{} and Jinniu Hu *\orcidB{}}

\AuthorNames{Firstname Lastname, Firstname Lastname and Firstname Lastname}

{%
	\isChicagoStyle{%
		\AuthorCitation{Firstname, Firstname, Firstname Lastname, and Firstname Lastname.}
	}{
		\AuthorCitation{Li, S.; Shen, H.; Hu, J.}
	}
}

\address[1]{School of Physics, Nankai University, Tianjin 300071, China; 1120240084@mail.nankai.edu.cn (S.L.)}

\corres{\hangafter=1 \hangindent=1.05em \hspace{-0.82em}Correspondence: shennankai@gmail.com (H.S.); hujinniu@nankai.edu.cn (J.H.)}

\abstract{The Thomas--Fermi approximation is a powerful method that has been widely used to describe 
	atomic structures, finite nuclei, and nonuniform matter in supernovae and neutron-star crusts.
	Nonuniform nuclear matter at subnuclear density is assumed to be composed of a lattice of heavy 
	nuclei surrounded by dripped nucleons, and the Wigner--Seitz cell is commonly introduced to simplify 
	the calculations. The self-consistent Thomas--Fermi approximation can be employed to study both a 
	nucleus surrounded by nucleon gas in the Wigner--Seitz cell and an isolated nucleus in the nuclide chart.
	A detailed comparison is made between the self-consistent Thomas--Fermi approximation and the 
	relativistic mean-field approach for the description of finite nuclei, based on the same nuclear 
	interaction. These results are then examined using experimental data from the corresponding nuclei.}

\begin{document}
	
	\section{Introduction}
	\label{Introduction}
	
	The Thomas--Fermi approximation has a long history in the study of atomic structures, finite nuclei, 
	and nonuniform matter appearing in astrophysical environments.
	In its early development, the Thomas--Fermi approximation for finite nuclei was typically employed 
	within nonrelativistic theoretical frameworks~\cite{Bethe1968, BRACK1985275, CENTELLES1990397}. 
	In 1977, Boguta and Rafelski~\cite{BOGUTA197722} combined the Thomas--Fermi approximation with a
	relativistic Hartree approach, enabling investigations of finite nuclear densities. 
	In Boguta and Bodmer~\cite{boguta1977nuclear_matter}, the authors applied both a relativistic Hartree 
	approach and Thomas--Fermi approximation to study the nuclear surface properties of semi-infinite nuclear matter. 
	Both approaches yielded remarkably similar average nuclear properties. 
	Their studies revealed that, within the Thomas--Fermi approximation for finite nuclei, 
	the average surface properties were consistent with those for semi-infinite nuclear matter.
	This consistency motivated the combination of Thomas--Fermi approximation and a relativistic Hartree 
	approach for the study of finite nuclei. 
	Walecka et~al.~\cite{SERR197810, SEROT1979172} further integrated the Thomas--Fermi approximation 
	within a covariant density functional theory, discussing the selection of model parameters and 
	performing detailed calculations for $^{40}\text{Ca}$ and $^{208}\text{Pb}$. 
	The Thomas--Fermi approximation has also been used to study hot nuclei. 
	Zhang et al.~\cite{Zhang_Bao_2014} developed a description of hot nuclei within a self-consistent 
	Thomas--Fermi (STF) approximation using the covariant density functional theory for nuclear interactions. 
	They used a subtraction procedure to isolate the nucleus from nucleon gas inside a Wigner--Seitz cell,
	and then calculated the temperature dependence of nuclear symmetry energy. 
	The surface energy and nucleon distribution are determined in a self-consistent manner
	within the STF approximation by solving the relevant coupled~equations.
	
	In recent years, the Thomas--Fermi approximation combined with 
	\textcolor{black}{covariant density functional theory} has been applied to 
	calculate the equation of state (EOS) for nonuniform nuclear matter existing in supernovae and 
	neutron-star crusts, owing to its computational simplicity and 
	effectiveness~\cite{Avancini2008, Zhang_2014, Oertel_2017}. 
	The nonuniform nuclear matter was assumed to be composed of a lattice of heavy nuclei, surrounded
	by free nucleons, while the Wigner--Seitz cell is commonly introduced to simplify the calculations. 
	One of the most commonly used EOS in astrophysical simulations, often referred to as Shen 
	EOS~\cite{Shen_1998, Shen_2011, Shen_2020}, employs the Thomas--Fermi approximation with 
	parameterized nucleon distributions inside the Wigner--Seitz cell. 
	At low temperatures and subnuclear densities, complex inhomogeneous matter, known as nuclear pasta phases,
	may appear, which is likely to occur in neutron-star crusts. 
	Avancini et al.~\cite{Avancini_2009, Avancini_2010} employed the STF approximation to study 
	the properties of pasta phases. Their studies showed that the STF approximation can describe
	not only an isolated nucleus but also a nucleus surrounded by nucleon gas within the Wigner--Seitz cell.
	
	\textcolor{black}{The relativistic quantum field theory for nuclear many-body systems, proposed by 
		Walecka et al.~\cite{SERR197810, SEROT1979172}, is nowadays considered as a covariant form 
		of density functional theory. Over the past few decades, the covariant density functional theory 
		has undergone significant development and achieved great success in studies on various nuclear 
		phenomena~\cite{RING1996193, Rufa_1988,Avancini2008,Oertel_2017}. 
		This theory establishes a complete relativistic interaction system by introducing scalar and vector 
		mesons as mediators of nucleon--nucleon interactions. In this work, we refer to its application to
		finite nuclei as the relativistic mean-field (RMF) approach.}
	The RMF approach has been successfully applied to study the ground-state properties of finite nuclei, 
	such as binding energies, density distributions, and single-particle spectra~\cite{reinhard1986}. 
	By incorporating proper treatment of pairing correlations, the RMF approach can be systematically 
	extended to investigate nuclei across the nuclide chart, which reflects its great capability for 
	reproducing experimental observations.
	
	As two fundamental approaches for describing finite nuclei, the quantitative differences 
	between the STF approximation and the RMF approach deserve to be examined.
	Within the STF approximation, the nucleon field operators are expanded on a plane-wave basis, 
	treating the nucleon system as a relativistic Fermi gas~\cite{SERR197810, Von-Eiff_1797, Zhang_2014}. 
	This approximation requires only self-consistent solutions of meson field equations to obtain 
	the source distributions directly, without the need to explicitly handle the Dirac spinor of nucleons. 
	By employing this semi-classical treatment of nucleons, it avoids the complicated calculations 
	involved in solving the Dirac equation in a central potential, which is particularly suitable 
	for studying heavy nuclei within nonuniform matter existing in astrophysical contexts.
	In contrast, the RMF approach adopts a more microscopic method by solving the Dirac equation of nucleons
	together with the meson field equations in a self-consistent manner.
	The nucleon field can be explicitly decomposed into radial and angular components, 
	with nucleon energy levels obtained by solving the radial Dirac equation~\cite{GAMBHIR1990132,RING1996193}. 
	This method accurately provides important physical quantities such as the intensity of spin-orbit 
	interactions and single-nucleon spectra~\cite{ebran2016}, making it particularly suitable for 
	detailed studies of finite nuclei. 
	\textcolor{black}{Generally, the parameter set in the RMF approach is determined by fitting to 
		the saturation properties of infinite nuclear matter and/or selected experimental data of finite nuclei.}
	In this work, we first employ the STF approximation to calculate various properties of finite nuclei using several typical RMF parameter sets, and then conduct a detailed comparison between the STF approximation and the RMF approach for describing finite nuclei.
	
	The paper is organized as follows. In Section~\ref{Materials and Methods}, we briefly introduce 
	the theoretical framework for finite nuclei, including the STF approximation and RMF approach.
	Furthermore, we provide a new parameter set, TM1m*, and compare it with other parameter sets. 
	In Section~\ref{Results and Discussion}, we present the numerical results for finite nuclei 
	using several parameter sets. Additionally, we conduct a detailed comparison between the STF approximation 
	and the RMF approach for describing finite nuclei. Section~\ref{Conclusions} is devoted to the conclusions.
	
	\section{Materials and Methods}
	\label{Materials and Methods}
	
	We first provide a brief description of the \textcolor{black}{covariant density functional theory} used for 
	a nuclear many-body system. 
	Subsequently, the details of the STF approximation and the RMF approach for finite nuclei are 
	discussed in two separate subsections.
	
	In the \textcolor{black}{covariant density functional theory}, nucleons interact via the exchange of 
	isoscalar scalar and vector mesons ($\sigma$ and $\omega$) and isovector vector meson ($\rho$). 
	When describing a finite nuclear system, the electromagnetic field ($A^{\mu}$) cannot be neglected. 
	We employ the \textcolor{black}{covariant density functional theory}, including nonlinear terms for 
	the $\sigma$ and $\omega$ mesons, as well as an $\omega$-$\rho$ coupling term~\cite{Bao_2014,Zhang_2014}. 
	The nonlinear terms for the $\sigma$ meson address the issue of the excessive incompressibility of 
	nuclear matter~\cite{boguta1977nuclear_matter}, while the addition of the nonlinear term for the $\omega$ meson 
	reproduces the density-dependent nucleon self-energy obtained from the relativistic 
	Brueckner--Hartree--Fock theory~\cite{SUGAHARA1994557}. The $\omega$-$\rho$ coupling term is 
	introduced to modify the density dependence of nuclear symmetry energy~\cite{Fattoyev_2010}. 
	The Lagrangian density of the \textcolor{black}{covariant density functional theory} takes the form
	\vspace{-12pt}
	\begin{adjustwidth}{-\extralength}{0cm}
		\centering 
		\begin{eqnarray}
			\label{lagrangian density}
			{\cal L} &=& \sum\limits_{i=p,n}\bar{\psi}_{i}
			\left\{i\gamma_{\mu}\partial^{\mu}-(M+g_{\sigma}\sigma)
			-\gamma_{\mu}\left[g_{\omega}\omega^{\mu}+\frac{g_{\rho}}{2}\tau_{a}\rho^{a\mu}
			+\frac{e}{2}(1+\tau_3)A^{\mu}\right]\right\}\psi_{i} \nonumber \\
			& & +\frac{1}{2}\partial_{\mu}\sigma\partial^{\mu}\sigma-\frac{1}{2}m_{\sigma}^{2}\sigma^{2}
			-\frac{1}{3}g_{2}\sigma^{3}-\frac{1}{4}g_{3}\sigma^{4}  -\frac{1}{4}W_{\mu\nu}W^{\mu\nu}+\frac{1}{2}m_{\omega}^{2}\omega_{\mu}\omega^{\mu}
			+\frac{1}{4}c_{3}(\omega_{\mu}\omega^{\mu})^{2}  \nonumber \\
			& & -\frac{1}{4}R^{a}_{\mu\nu}R^{a\mu\nu}+\frac{1}{2}m_{\rho}^{2}\rho^{a}_{\mu}\rho^{a\mu}
			+\Lambda_{\text{v}}(g_{\omega}^{2}\omega_{\mu}\omega^{\mu})(g_{\rho}^{2}\rho_{\mu}^{a}\rho^{a\mu})
			-\frac{1}{4}F_{\mu\nu}F^{\mu\nu},
		\end{eqnarray}
	\end{adjustwidth}
	where $W^{\mu\nu}$, $R^{a\mu\nu}$, $F^{\mu\nu}$ represent the antisymmetric field tensors for $\omega^{\mu}$, $\rho^{a\mu}$, $A^{\mu}$, respectively. In the mean-field approximation, the meson fields are treated as classical fields and the meson field operators are replaced by their expectation values, namely, 
	$\sigma = \braket{\sigma}$, \mbox{$\omega = \braket{\omega^{0}}$}, $\rho = \braket{\rho^{30}}$. 
	The equations of motion for these mesons, derived from the Lagrangian 
	density,~(\ref{lagrangian density}) are written as
	\begin{eqnarray}
		\label{sigma field}
		&& -\nabla^{2}\sigma+m_\sigma^2 \sigma+g_2 \sigma^2+g_3 \sigma^3 = -g_\sigma \left(n^{s}_{p}+n^{s}_{n}\right), \\
		\label{omega field}
		&& -\nabla^{2}\omega+m_\omega^2 \omega+c_3\omega^{3}+2\Lambda_{\text{v}}(g_{\omega}^{2}\omega)(g_{\rho}^{2}\rho^{2}) = g_\omega \left(n_{p}+n_{n}\right), \\
		\label{rho field}
		&& -\nabla^{2}\rho+m_\rho^2 \rho+2\Lambda_{\text{v}}(g_{\omega}^{2}\omega^{2})(g_{\rho}^{2}\rho) = \frac{g_\rho}{2} \left(n_{p}-n_{n}\right), \\
		\label{ele field}
		&& -\nabla^{2}A = e n_{p},
	\end{eqnarray}
	where $n^{s}_{i}$ and $n_{i}$ denote, respectively, the scalar density and vector density of species $i$. 
	In a finite nuclear system, the meson fields are assumed to exhibit spherically symmetric spatial profiles. 
	This assumption allows for the existence of spatial derivatives of these meson fields, 
	and quantities such as densities and mean fields are dependent on the radial position~$r$.
	
	\subsection{Self-Consistent Thomas--Fermi Approximation for Finite Nuclei}
	\label{subsection:STF}
	
	In the STF approximation, the variations of the meson fields are sufficiently slow, 
	which allows baryons to be treated as moving in locally constant fields~\cite{SEROT1979172}. 
	Consequently, the source terms of meson fields can be considered to have local spatial dependence. 
	The local scalar and vector densities are given by
	
	\begin{eqnarray}
		\label{scalar densities}
		&& n^{s}_{i} \left(r\right) = \frac{1}{\pi^{2}} \int_{0}^{k_{i}^{F}} 
		\frac{M^{\ast}}{\sqrt{k^{2}+M^{\ast 2}}} k^{2} ~dk,\\
		\label{vector densities}
		&& n_{i} \left(r\right)= \frac{1}{\pi^{2}} \int_{0}^{k_{i}^{F}} k^{2} ~dk
		= \frac{\left[{k_{i}^{F}\left(r\right)}\right]^3}{3\pi^{2}},
	\end{eqnarray}
	where $k_{i}^{F}$ is the Fermi momentum of species $i$ and $M^{\ast} = M+g_{\sigma}\sigma$ is the effective nucleon mass. The numbers of protons and neutrons inside a nucleus can be determined from the local 
	vector densities,
	\begin{eqnarray}
		&& N_{p} = \int_{\text{nucleus}} n_{p}\left(r\right) ~d^{3}r, \\
		&& N_{n} = \int_{\text{nucleus}} n_{n}\left(r\right) ~d^{3}r.
	\end{eqnarray}
	
	The baryon number conservation implies that chemical potentials of protons and neutrons 
	are spatially constant throughout the nucleus, which are given by
	\begin{eqnarray}
		&& \mu_p = \sqrt{\left(k_p\right)^2+M^{* 2}}+g_\omega \omega+\frac{g_\rho}{2} \rho+e A, \\
		&& \mu_n = \sqrt{\left(k_n\right)^2+M^{* 2}}+g_\omega \omega-\frac{g_\rho}{2} \rho.
	\end{eqnarray}
	
	In the STF approximation for heavy nuclei, the center-of-mass corrections and the pairing correlations 
	can be disregarded, as previously discussed in nonrelativistic Thomas--Fermi calculations~\cite{Von-Eiff_1797}.
	Consequently, the total energy of a nucleus is given by
	\begin{equation}
		\label{STF energy}
		\begin{aligned}
			E_{\text{total}}=E_{\text{STF}} = 
			& \int d^{3} r~\left[~\sum_{i=p,n} \frac{1}{\pi^{2}} 
			\int_{0}^{k_{i}^{F}} dk~k^{2} \sqrt{k^{2}+M^{\ast 2}} \right. \\
			& -\frac{1}{2} g_\sigma \sigma \left(n_{p}^{s}+n_{n}^{s}\right)+\frac{1}{2} g_\omega \omega\left(n_p+n_n\right)+\frac{1}{4} g_\rho \rho\left(n_p-n_n\right)+\frac{1}{2} e A n_p \\
			& \left. -\frac{1}{6} g_2 \sigma^3-\frac{1}{4} g_3 \sigma^4 + \frac{1}{4} c_3 \omega^4+\Lambda_{\text{v}}\left(g_\omega^2 \omega^2\right)\left(g_\rho^2 \rho^2\right) \right],
		\end{aligned}
	\end{equation}
	where Equations~(\ref{sigma field})--(\ref{ele field}) are used to simplify 
	the calculations of the derivatives of mean~fields.
	
	In practical calculations, we first specify the number of protons and neutrons, $N_{p}$ and $N_{n}$,
	in the nucleus. Then, we provide initial guesses for the meson fields $\sigma \left(r\right)$, $\omega \left(r\right)$, $\rho \left(r\right)$, and the electromagnetic field $A \left(r\right)$. 
	These initial guesses are used to determine the chemical potentials, $\mu_p$ and $\mu_n$, 
	subject to the constraints $N_{p}$ and $N_{n}$. Once the chemical potentials are achieved, 
	we can calculate various densities using Equations~(\ref{scalar densities}) 
	and (\ref{vector densities}), which will be used to solve Equations~(\ref{sigma field})--(\ref{ele field}) 
	to obtain new mean fields. If the resulting fields differ from the initial guesses, 
	the guesses are updated and the processes are repeated until the fields converge to the guesses.
	
	\subsection{Relativistic Mean-Field Approach for Finite Nuclei}
	\label{subsection:RMF}
	
	\textcolor{black}{In the covariant density functional theory, no-sea approximation is imposed.} 
	The nucleon field operator can be expanded in terms of single-particle wave functions, 
	which is written as
	\begin{eqnarray}
		\psi_{i} = \left[\sum_{\alpha} A_{\alpha} \phi_{\alpha} \right]_{i},
	\end{eqnarray}
	where $A_{\alpha}$ denotes the annihilation operator for nucleons and $\phi_{\alpha}$ represents the  
	single-particle wave function, with $\alpha$ being a set of quantum numbers. 
	For a spherically symmetric nuclear system, the set of quantum numbers, $\alpha$, includes the 
	conventional quantum numbers for angular momentum and parity. 
	The nucleon wave function $\phi_{\alpha}$ can then be divided into radial and angular parts,
	\begin{eqnarray}
		\label{wavefun}
		\phi_{\alpha}=\left(\begin{array}{c}
			\mathrm{i}\left[G_{n \kappa}(r) / r\right] \Phi_{\kappa m} \\
			-\left[F_{n \kappa}(r) / r\right] \Phi_{-\kappa m}
		\end{array}\right),
	\end{eqnarray}
	where $n$ is the principal quantum number, $\kappa$ is determined by the angular momentum quantum 
	numbers $j$ and $l$, and $m$ is the magnetic quantum number. The upper and lower components have 
	opposite values of $\kappa$, and they are normalized 
	by $\int_{0}^{\infty} \left( \lvert G_{\alpha}\left(r\right) \rvert ^{2} + \lvert F_{\alpha}\left(r\right) \rvert^{2} \right) dr= 1$. They satisfy the coupled equations
	\begin{equation}
		\label{wavefun_G}
		\small
		\frac{d }{d r}G_{\alpha}\left(r\right)+\frac{k}{r}G_{\alpha}\left(r\right)=M^{*}F_{\alpha}\left(r\right)
		-\left[g_\omega \omega+\frac{g_\rho}{2} \tau_3\rho+\frac{e}{2}(1+\tau_3)A\right]
		F_{\alpha}\left(r\right)+E_{\alpha} F_{\alpha}\left(r\right),
	\end{equation}
	\begin{equation}
		\label{wavefun_F}
		\small
		\frac{d }{d r}F_{\alpha}\left(r\right)-\frac{k}{r}F_{\alpha}\left(r\right)=M^{*}G_{\alpha}\left(r\right)
		+\left[g_\omega \omega+\frac{g_\rho}{2} \tau_3\rho+\frac{e}{2}(1+\tau_3)A\right]
		G_{\alpha}\left(r\right)-E_{\alpha} G_{\alpha}\left(r\right).
	\end{equation}
	
	When the nucleon wave function $\phi_{\alpha}$ is obtained, we can calculate various densities in
	the source terms of meson field equations, such as
	\begin{eqnarray}
		&& n_{i}^{s} \left(r\right) = \left[\sum_{\alpha}^{\text{occ}} w_{\alpha} \bar{\phi}_{\alpha} \phi_{\alpha}\right]_{i} = \left[\sum_{\alpha}^{\text{occ}} w_{\alpha} \frac{2j_{\alpha}+1}{4 \pi r^{2}} \left( \lvert G_{\alpha} \left(r\right) \rvert^{2} - \lvert F_{\alpha} \left(r\right) \rvert^{2}\right)\right]_{i} , \\
		\label{density_RMF}
		&& n_{i} \left(r\right) = \left[\sum_{\alpha}^{\text{occ}} w_{\alpha} \bar{\phi}_{\alpha} \gamma_{0} \phi_{\alpha}\right]_{i} = \left[\sum_{\alpha}^{\text{occ}} w_{\alpha} \frac{2j_{\alpha}+1}{4 \pi r^{2}} \left( \lvert G_{\alpha} \left(r\right) \rvert^{2} + \lvert F_{\alpha} \left(r\right) \rvert^{2}\right)\right]_{i}.
	\end{eqnarray}
	
	Here, we incorporate the pairing correlations based on the BCS theory to extend the study beyond
	closed-shell nuclei~\cite{reinhard1986, GAMBHIR1990132}, introducing the occupation probability $w_{\alpha}$ 
	for each nucleon state. When the number of protons or neutrons corresponds to a magic number, 
	the occupation probability equals one for occupied states and zero for unoccupied states.
	Otherwise, the occupation probability is evaluated by
	\begin{eqnarray}
		w_{\alpha} = \frac{1}{2} \left(1-\frac{E_{\alpha}-\lambda}{\sqrt{\left(E_{\alpha}-\lambda\right)^{2}+\Delta^{2}}}\right),
	\end{eqnarray}
	where $E_{\alpha}$ is the single-particle energy for a quantum state. 
	The gap energy $\Delta$ is taken to be
	$\Delta = 11.2~\text{MeV}/\left(N_{p}+N_{n}\right)^{1/2}$, as given in~\cite{SUGAHARA1994557}. 
	The Fermi energy $\lambda$ is determined by the condition
	\begin{eqnarray}
		\left[\sum_{\alpha} w_{\alpha}\right]_{i}=N_{i}.
	\end{eqnarray}
	
	The energy correction arising from pairing correlations is given by
	\begin{eqnarray}
		E_{\text{pair}} = -\Delta \sum_{i=p, n} \sum_{\alpha} \left[\sqrt{w_{\alpha} \left(1-w_{\alpha}\right)}\right]_{i}.
	\end{eqnarray}
	
	In addition, we need to make the center-of-mass corrections on the total energy. 
	For the sake of computational convenience, we employ the center-of-mass correction derived from the nonrelativistic harmonic oscillator potential, which is expressed as
	\begin{eqnarray}
		E_{\text{ZPE}} = \frac{3}{4} \times 41 \left(N_{p}+N_{n}\right)^{-1/3} \ ~\left(\text{MeV}\right). 
	\end{eqnarray}
	
	To ensure completeness, we give an expression for the total energy of the system,
	\begin{eqnarray}
		E_{\text{total}}=E_{\text{RMF}}+E_{\text{pair}}-E_{\text{ZPE}},
	\end{eqnarray}
	where $E_{\text{RMF}}$ is calculated in the RMF approach as
	\begin{equation}
		\begin{aligned}
			E_{\text{RMF}}= & \sum_{i=p, n} \sum_{\alpha}^{\text{occ}} \left[w_{\alpha} \left(2j_{\alpha}+1\right)E_{\alpha}\right]_{i} \\
			& +\int d^{3} r~\left[-\frac{1}{2} g_\sigma \sigma \left(n_{p}^{s}+n_{n}^{s}\right) 
			-\frac{1}{2} g_\omega \omega\left(n_p+n_n\right)  
			-\frac{1}{4} g_\rho \rho\left(n_p-n_n\right)-\frac{1}{2} e A n_p\right. \\
			& \left. -\frac{1}{6} g_2 \sigma^3-\frac{1}{4} g_3 \sigma^4+\frac{1}{4} c_3 \omega^4
			+\Lambda_{\text{v}}\left(g_\omega^2 \omega^2\right)\left(g_\rho^2 \rho^2\right) \right].
		\end{aligned}
	\end{equation}
	
	\textcolor{black}{The total energy expression in the RMF approach incorporates baryon field 
		contributions calculated by solving the Dirac equation, whereas the Thomas--Fermi approximation 
		employs momentum integration combined with source terms from Equations~(\ref{sigma field})--(\ref{ele field}). 
		This fundamental methodological difference ultimately manifests as discrepancies in the resulting
		total energies.}
	
	\subsection{Parameter Sets}
	\label{subsection:Pars}
	
	Recent studies by Schneider et al.~\cite{Schneider_2019} and Yasin et al.~\cite{Yasin_2020} have shown 
	that a larger effective mass provides a more favorable environment for the shock evolution and explosion 
	of core-collapse supernova. However, their studies are limited to nonrelativistic Skyrme models. 
	\textcolor{black}{To investigate the effects of the effective nucleon mass within a relativistic framework, we refitted the TM1e ~\cite{Bao_2014} parameter set in the RMF approach, which is referred to as the TM1m model in our previous work~\cite{Li_2025}. }
	The new TM1m model features a high effective mass ratio $M^{\ast}/M \sim 0.8$ at saturation density, 
	while maintaining the same saturation properties as the TM1e model.
	Using the TM1m model, we constructed a new equation of state (EOS5) for uniform nuclear matter 
	at densities higher than $\sim$$10^{14}\, \text{g}/\text{cm}^{3}$, which is applicable to 
	complex astrophysical phenomena, such as core-collapse supernovae and binary neutron star mergers.
	
	At subsaturation densities and low temperatures, nucleons may form nonuniform matter including heavy nuclei 
	and light clusters to lower the free energy of the system. To investigate nonuniform matter, 
	both liquid-drop models and Thomas--Fermi approximations are widely employed. In these models, 
	the parameters are relevant to their predictions for the ground-state properties of finite nuclei. 
	However, the results of the TM1m model for finite nuclei are not in good agreement with 
	relevant experimental data.
	In the present work, we incorporate the RMF calculations for finite nuclei, as described 
	in Section~\ref{subsection:RMF}, into the parameter adjustment process.
	We adjust the new parameter set to match the energies per nucleon and charge radii 
	of $^{90} \text{Zr}$, $^{116} \text{Sn}$, $^{124} \text{Sn}$, $^{196} \text{Pb}$, $^{208} \text{Pb}$, $^{214} \text{Pb}$~\cite{Wang_2021, ANGELI201369}. The energy per nucleon is given by
	\begin{eqnarray}
		E/A = E_{\text{total}}/A - M,
	\end{eqnarray}
	where $A=N_p+N_n$ denotes the mass number of the nucleus. 
	The adopted charge radius is calculated using the formula given in~\cite{SUGAHARA1994557}:
	\begin{eqnarray}
		\label{RC}
		R_{c}^{2} = \left[R_{p}^{2}+\left(0.862~\text{fm}\right)^{2}-\left(0.336~\text{fm}\right)^{2} \frac{N_{n}}{N_{p}}\right].
	\end{eqnarray}
	
	We obtain a new parameter set, referred to as TM1m*, which features a large effective mass 
	and improved agreement with the ground-state properties of finite nuclei.
	
	The parameter sets of TM1m*, TM1m, TM1e, and their original TM1 models are presented in Table~\ref{parameter}, 
	while the corresponding saturation properties are listed in Table~\ref{properties}. 
	\textcolor{black}{The original TM1 model has a rather large slope parameter of the symmetry 
		energy (\mbox{$L=111$ MeV}), which predicts too large radii for neutron stars as compared to the 
		estimations from astrophysical observations~\cite{Shen_2020}. In contrast, a small value of
		$L=40$ MeV has been achieved in the TM1e, TM1m, and TM1m* models by introducing an additional
		$\omega$-$\rho$ coupling term. The main difference between TM1e and TM1m lies in 
		the effective mass, with TM1e having $M^{\ast}/M \sim 0.63$ and TM1m having $M^{\ast}/M \sim 0.8$.
		Both TM1m and TM1m* have relatively large effective masses, whereas their saturation properties 
		are almost identical. The difference between TM1m and TM1m* is attributed to distinct fitting procedures.
		Specifically, the ground-state properties of finite nuclei are incorporated into the fitting process 
		for the TM1m* parameter set to improve its description of finite nuclei.
		In contrast, only saturation properties of nuclear matter were considered in the fitting process 
		for the TM1m parameter set. Therefore, the TM1m* model is able to provide better descriptions for 
		finite nuclei than the TM1m model.
		Notably, the effective mass ratio $M^{\ast}/M \sim 0.8$ in TM1m and TM1m* is an extreme 
		large value within the covariant density functional theory~\cite{Li_2025}.}
	In addition, we consider two well-known 
	RMF models, NL3~\cite{Lalazissis_1997} and SFHo~\cite{Steiner_2013}, in our calculations for comparison. 
	As shown in Table~\ref{properties}, the saturation properties of NL3 are similar to those 
	of the TM1 model, particularly in terms of the symmetry energy and its slope. 
	The NL3 model has been shown to perform well in describing nuclear properties~\cite{LALAZISSIS_1999, Priyanka_2024}. The SFHo model is commonly used in astrophysical simulations~\cite{Figura_2020, Boukari_2021}. 
	It is characterized by multiple coupling terms for the $\rho$ meson, and its effective mass 
	ratio $M^{\ast}/M \sim 0.76 $ is comparable to that of the TM1m and TM1m* models.
	\textcolor{black}
	{There are alternative RMF models, such as the DDRMF models with density-dependent 
		couplings~\cite{typel1999,typel2020}, which have proven to be very powerful in describing
		nuclear matter and finite nuclei. Furthermore, the DDRMF model including tensor couplings can increase 
		the effective nucleon masses to be about $M^{\ast}/M \sim 0.67$~\cite{typel2020}. 
		For simplicity, we do not consider the DDRMF models in the present calculations.}
	
	\begin{table}[H]
		\caption{Parameter sets of TM1m*, TM1m, TM1e, TM1, and NL3 models.}
		\label{parameter}
		\small
		\begin{tabularx}{\textwidth}{CCCCCC}
			\toprule	
			& \textbf{TM1m*} & \textbf{TM1m} & \textbf{TM1e} & \textbf{TM1} & \textbf{NL3}\\
			\midrule
			$m_{\sigma}$ (MeV) & 463.680 & 511.198 & 511.198 & 511.198 & 508.194 \\
			$g_{\sigma}$  & 7.15454 & 7.9353 & 10.0289 & 10.0289 & 10.217  \\
			$g_{\omega}$ & 8.59221 & 8.6317 & 12.6139 & 12.6139 & 12.868 \\
			$g_{\rho}$ & 11.52164 &11.5130 &13.9714  & 9.2644 & 8.948 \\
			$g_{2}$ ($\text{fm}^{-1}$) & $-$7.86677 & $-$11.5163  & $-$7.2325 & $-$7.2325 & $-$10.431 \\
			$g_{3}$  & 40.55692 & 54.8872 & 0.6183 & 0.6183 & $-$28.885 \\
			$c_{3}$  & 0 & 0.00025 & 71.3075 & 71.3075  & 0 \\
			$\Lambda_{\text{v}}$  & 0.09465 & 0.0933 & 0.0429 & 0 & 0  \\
			\bottomrule
		\end{tabularx}
	\end{table}
	\unskip
	\begin{table}[H]
		\caption{Nuclear matter properties of TM1m*, TM1m, TM1e, TM1, NL3, and SFHo models. 
			The saturation density and energy per particle are denoted by $n_{0}$ and $E/A$, respectively. 
			The incompressibility is represented by $K$, the symmetry energy by $E_{\text{sym}}$, 
			the slope of the symmetry energy by $L$, and the effective mass ratio by $ M^{\ast}/M $.}
		\label{properties}
		\begin{tabularx}{\textwidth}{CCCCCCC}
			\toprule	
			& \textbf{TM1m*} & \textbf{TM1m} & \textbf{TM1e} & \textbf{TM1} & \textbf{NL3} &\textbf{SFHo}   \\
			\midrule
			$n_0\,(\text{fm}^{-3})$ & 0.145&0.145&0.145&0.145 & 0.148 &0.158\\
			$E/A\,\text{(MeV)}$   & $-$16.2 & $-$16.3&$-$16.3&$-$16.3&$-$16.2 & $-$16.2\\
			$K\,\text{(MeV)}$  &282 & 281&281&281&272 & 245\\
			$E_\text{sym}\,\text{(MeV)}$  &31.4&31.4&31.4 & 36.9&37.4 & 31.6 \\
			$L\,\text{(MeV)}$ &40 &40 &40 &111 & 118 & 47 \\
			$M^{*}/M$  & 0.794 &0.793 & 0.634& 0.634 & 0.595 & 0.761 \\
			\bottomrule
		\end{tabularx}
	\end{table}
	
\begin{figure}[H]
		\includegraphics[width=0.7\linewidth]{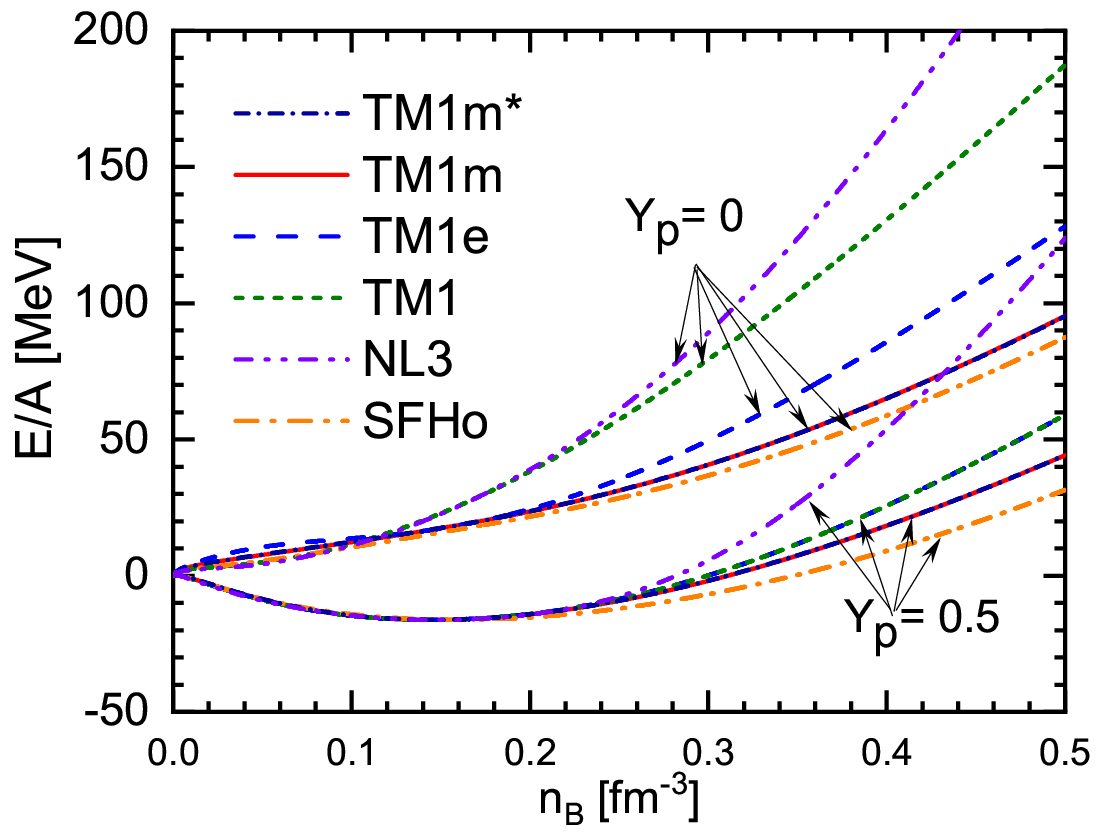}
		\caption{Energy per nucleon $E/A$ of symmetric nuclear matter and pure neutron matter as 
			a function of baryon number density $n_{B}$.}
		\label{21EA}
	\end{figure}

	In Figure~\ref{21EA}, we display the energy per nucleon of symmetric nuclear matter and neutron matter 
	as a function of baryon number density $n_{B}$. It is shown that both the TM1m* and TM1m models 
	exhibit consistent behavior of symmetric nuclear matter, as do the TM1e and TM1 models. 
	The curves of TM1m* and TM1m for symmetric nuclear matter are slightly lower than those of TM1e and TM1, 
	which is because the larger effective masses in the TM1m* and TM1m models lead to smaller kinetic 
	energy contributions~\cite{Li_2025}. In pure neutron matter, the TM1m* and TM1m models still exhibit 
	identical behavior, as they possess the same symmetry energy characteristics.
	In contrast, the TM1e and TM1 models present different behaviors in pure neutron matter because of 
	their distinct symmetry energy slopes. The TM1 model shows a more rapid increase with increasing 
	density, while similar behavior is also observed in the NL3 model, due to their large slope values.
	At densities lower than $\sim$$0.2\, \text{fm}^{-3}$, the curves for symmetric nuclear matter and 
	pure neutron matter in the NL3 model are identical to those in the TM1 model, owing to their 
	nearly identical saturation properties. However, as the density increases, the NL3 curves rise rapidly. 
	This is because the vector potential in the NL3 model grows rapidly~\cite{DelEstal_2001, Long_2004}. 
	The TM1 model incorporates a quartic vector self-interaction, which reduces the vector potential 
	and yields relatively slower growth.  
	The curve of the SFHo model for symmetric nuclear matter lies below those of TM1m* and TM1m, 
	primarily because the SFHo model has a lower incompressibility than that of TM1m* and TM1m. 
	The value of the incompressibility directly affects the theoretical predictions of giant monopole 
	resonance (GMR) characteristics~\cite{Lalazissis_1997,Stone2014}.
	
\begin{figure}[H]
		\includegraphics[width=0.7\linewidth]{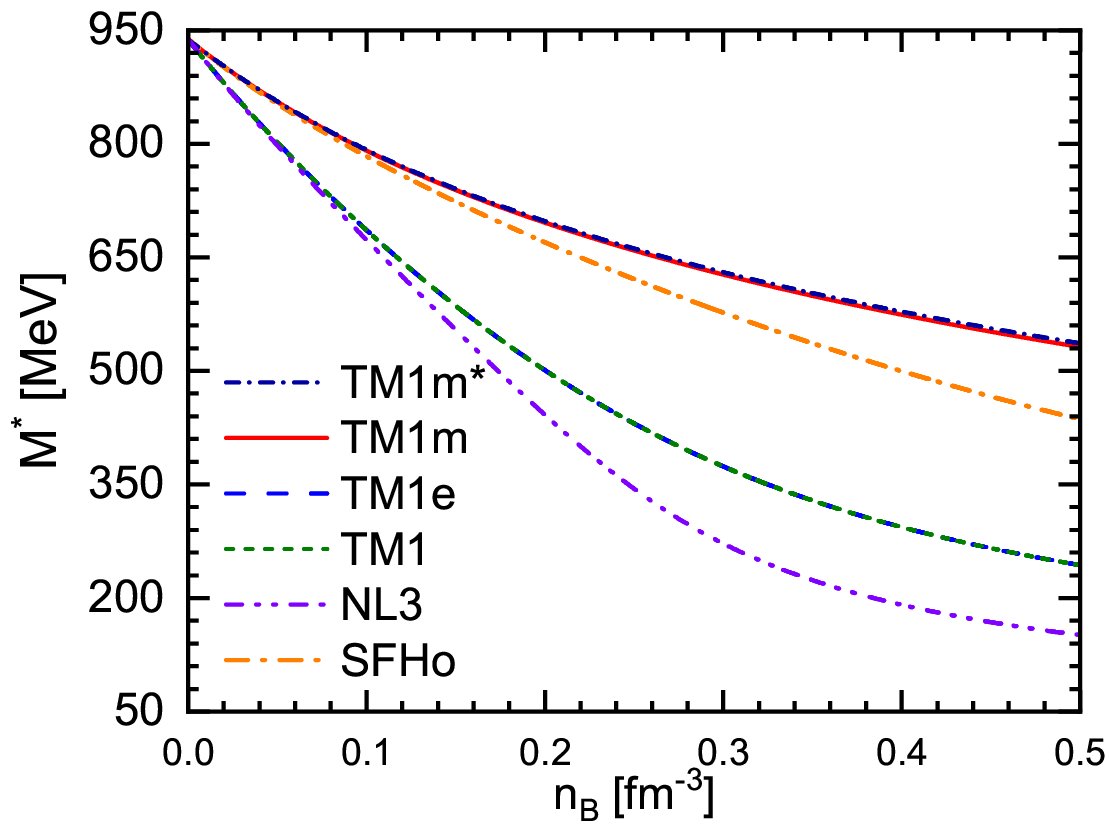}
		\caption{Effective nucleon mass $M^{\ast}$ as a function of baryon number density $n_{B}$. }
		\label{22Mn}
	\end{figure}

	In Figure~\ref{22Mn}, the effective nucleon mass $M^{\ast}$ is plotted as a function of the baryon 
	number density $n_{B}$. It is shown that the curves of the TM1e and TM1 models are identical due to 
	their same isoscalar properties. The curves of the TM1m* and TM1m models are significantly higher 
	than those of TM1e and TM1. The SFHo model exhibits a lower effective nucleon mass compared to 
	TM1m* and TM1m, but it is still much higher than that of TM1 and NL3. Notably, the NL3 model 
	exhibits the smallest effective nucleon mass among these parameterizations. 
	For finite nuclei, the effective mass influences the spin-orbit splitting~\cite{SHARMA1994110}, 
	which can be enhanced by introducing a tensor term into the model~\cite{liliani2016,typel2020}.
	
\begin{figure}[H]
		\includegraphics[width=0.7\linewidth]{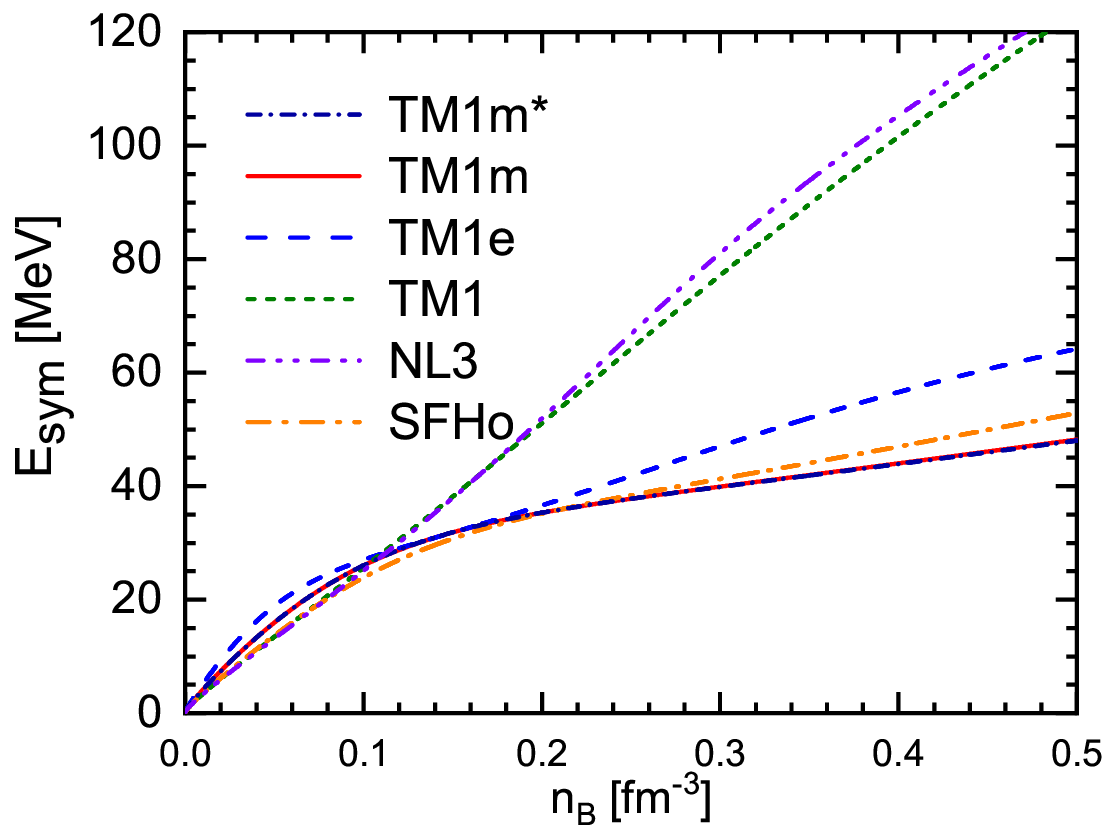}
		\caption{Symmetry energy $E_{\text{sym}}$ as a function of baryon number density $n_{B}$. }
		\label{23Esym}
	\end{figure}

	In Figure~\ref{23Esym}, we plot the symmetry energy $E_{\text{sym}}$ as a function of baryon number 
	density $n_{B}$. At high densities, the symmetry energy curves of the TM1 and NL3 models reach very 
	high values, while those of other models exhibit lower values. This difference is primarily due to 
	the distinct density dependence of the symmetry energy in these models. Specifically, the TM1 and NL3 
	models have significantly larger $L$ values than other models, as shown in Table~\ref{properties}. 
	This results in a more pronounced increase in symmetry energy at higher densities, which can 
	significantly influence the behavior in neutron-rich environments, such as neutron stars.
	On the other hand, the TM1m* and TM1m models exhibit lower symmetry energies than TM1e at high densities. Although the TM1m*, TM1m, and TM1e models have the same symmetry energy and slope values at saturation 
	density, the smaller effective mass and larger coupling parameters $g_{\sigma}$ and $g_{\omega}$ in
	the TM1e model favor more pronounced relativistic effects, which leads to a larger symmetry energy 
	at high densities. The SFHo model has a slightly larger symmetry energy slope compared to the TM1m* 
	and TM1m models, and therefore it predicts a larger symmetry energy than the TM1m* and TM1m models
	at high densities.
	
	\unskip
	
	\section{Results and Discussion} 
	\label{Results and Discussion}
	
	In this section, we systematically examine the ground-state properties for 19 doubly magic and semi-magic 
	nuclei that span a wide range of the nuclide chart, including $^{40}\text{Ca}$, $^{48}\text{Ca}$,
	$^{56}\text{Ni}$, $^{58}\text{Ni}$, $^{68}\text{Ni}$, $^{78}\text{Ni}$, 
	$^{90}\text{Zr}$, $^{130}\text{Cd}$, $^{100}\text{Sn}$, $^{116}\text{Sn}$, $^{124}\text{Sn}$, 
	$^{132}\text{Sn}$, $^{134}\text{Te}$, $^{144}\text{Sm}$, $ ^{182}\text{Pb}$, $^{184}\text{Pb}$, $^{196}\text{Pb}$, $^{208}\text{Pb}$, and $^{214}\text{Pb}$. 
	We exclude lighter nuclei because both the RMF approach and the STF approximation are inapplicable
	for few-body systems. In the present work, we employ six RMF parameter sets:, namely TM1m*, TM1m, 
	SFHo, TM1e, TM1, and NL3. Our analysis focuses on three fundamental nuclear properties, 
	specifically, (1) energy per nucleon, (2)~charge radius, and (3) nucleon density distribution.

	\subsection{Energy per Nucleon}
	\label{subsection:EA}
	
\begin{figure}[H]
		\begin{adjustwidth}{-\extralength}{0cm}
			\centering
			\includegraphics[width=1.0\linewidth]{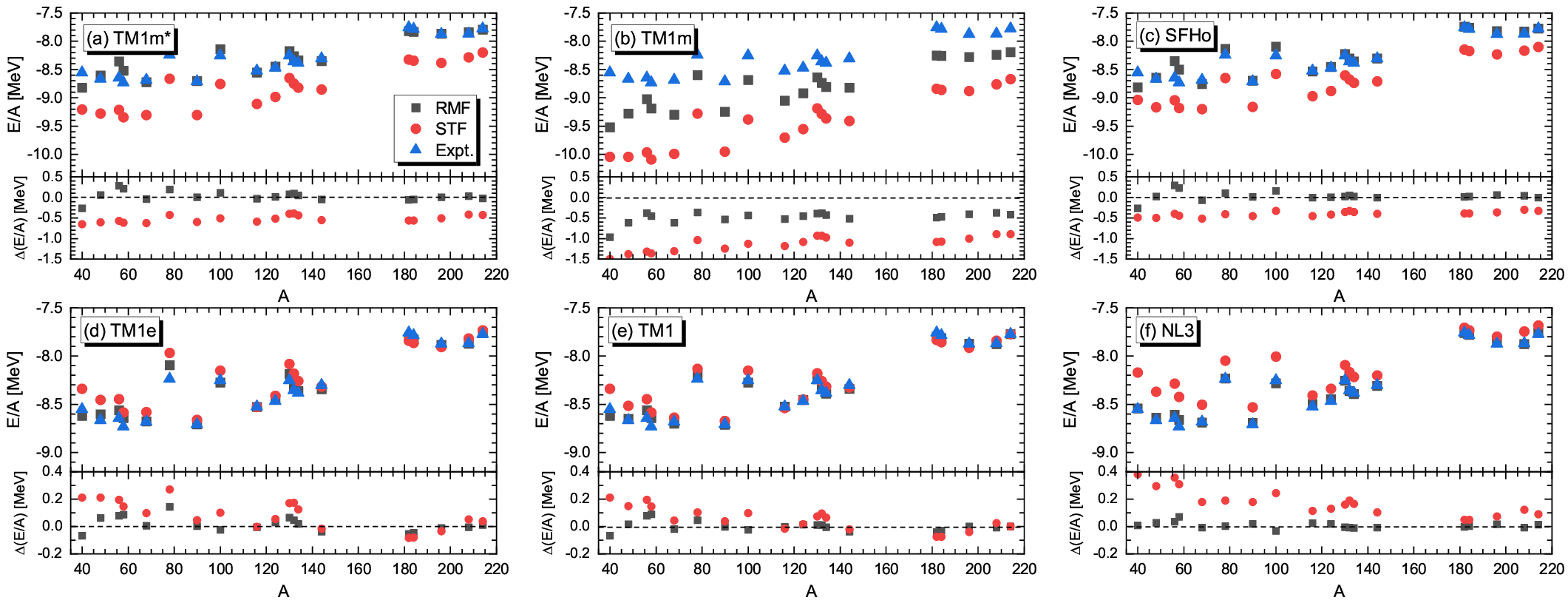}
		\end{adjustwidth}
	\caption{Energy per nucleon $E/A$ as a function of the mass number $A$ of a nucleus, 
		obtained in (\textbf{a})~TM1m*, (\textbf{b}) TM1m, (\textbf{c}) SFHo, (\textbf{d}) TM1e, (\textbf{e}) TM1, and (\textbf{f}) NL3. 
		The black squares and red circles correspond to the results obtained in the RMF approach and
		the STF approximation, respectively. The blue triangles represent experimental data 
		from AME2020~\cite{Wang_2021}. $\Delta (E/A)$ is the energy deviation between the model prediction
		and experimental data, defined as $\Delta (E/A)=(E/A)_{\text{RMF/STF}}-(E/A)_{\text{Expt.}}$. }
	\label{31BE}
\end{figure}

	In Figure~\ref{31BE}, we systematically compare the resulting energy per nucleon obtained in the RMF 
	approach and the STF approximation. The energy per nucleon $E/A$ for the 19 selected nuclei is presented 
	as a function of the mass number $A$, together with the experimental results, AME2020~\cite{Wang_2021}.
	To quantify the discrepancies between theoretical predictions and experimental values, we define 
	the energy deviation $\Delta (E/A)=(E/A)_{\text{RMF/STF}}-(E/A)_{\text{Expt.}}$, which is plotted 
	on the same $x$-axis together with the energy per nucleon $E/A$.
	Comparing Figure~\ref{31BE}a,b, it is shown that the deviations in the TM1m* model 
	are significantly smaller than those in the TM1m model. This is because the ground-state properties of 
	finite nuclei are included in the fitting process for the TM1m* parameter set, which helps to 
	improve the description of finite nuclei within the RMF approach. The results of the SFHo model, 
	as shown in Figure~\ref{31BE}c, are very similar to those of the TM1m* model, due to the comparable 
	fitting processes of both models. Comparing Figure~\ref{31BE}d,e, the results 
	obtained in the TM1e and TM1 models are very similar to each other, owing to their identical isoscalar 
	properties. Figure~\ref{31BE}f within the NL3 model also shows a similar pattern.
	Both the original TM1 and NL3 models employ the nuclear properties, such as the binding energies and charge
	radii, as inputs in the fitting procedures within the RMF framework~\cite{SHARMA1994110,Lalazissis_1997}, 
	and therefore their RMF results, shown by the black squares, are in excellent agreement with the empirical 
	values. Also, they have been shown to be successful in predicting various properties of finite 
	nuclei~\cite{Long_2002, REINHARD_2011, Rana_2022}. 
	However, within the STF approximation, the NL3 model predicts systematically higher values for the energy 
	per nucleon than the experimental values, indicating a tendency to overestimate the nuclear binding energy.
	In contrast, in the TM1m*, TM1m, and SFHo models, the STF results, shown by the red circles, fall 
	systematically below the experimental values. In the case of the TM1e and TM1 models, the STF results are 
	in good agreement with the RMF results shown by the black squares.
	Our analysis shows that the six models (TM1m*, TM1m, SFHo, TM1e, TM1, and NL3) exhibit varying degrees of discrepancies between the results from the RMF approach and those from the STF approximation.

	By comparing Figure~\ref{31BE}d,e, it is evident that the TM1e and TM1 models yield
	obviously different results for nuclei with relatively high isospin asymmetry, 
	such as $^{48}\text{Ca},\,^{78}\text{Ni},\,^{130}\text{Cd},\,^{132}\text{Sn}$, and $^{134}\text{Sm}$.
	It is found that the TM1e model predicts higher $E/A$ values for these nuclei compared to the TM1 model.
	This is because the density dependence of nuclear symmetry energy in the TM1e model differs from 
	that in the TM1 model, as shown in Figure~\ref{23Esym}. Compared to the TM1 model, the TM1e model has 
	a smaller symmetry energy slope, implying that the symmetry energy is smaller at supersaturation densities 
	and larger at subsaturation densities. This difference results in a more concentrated neutron distribution 
	and a more diffuse proton distribution at the nuclear surface~\cite{bao15,JiangWeiZhou_2010}. 
	Consequently, the overall nucleon distribution in nuclei with high isospin asymmetry becomes more 
	concentrated in the TM1e model than in the TM1 model. This implies that nucleons are more tightly 
	bound, thereby reducing the energy of the system. However, this effect is counteracted by increased 
	energy contributions from meson coupling terms, particularly the additional $\omega$-$\rho$ coupling 
	term in the TM1e model. Ultimately, the TM1e model yields a higher total energy than the TM1 model.

In order to analyze why the six models exhibit discrepancies between the results obtained using 
the RMF approach and those obtained using the STF approximation, we conduct a detailed comparison 
of the components of the energy. 
The total energy is decomposed into several components using Equation~(\ref{STF energy}), 
which includes the relativistic kinetic energy term; the density terms of the meson 
fields $\sigma$, $\omega$, and $\rho$ and the electromagnetic field $A$; the nonlinear terms of 
the $\sigma$ and $\omega$ mesons; and the $\omega$-$\rho$ coupling term. 
In Table~\ref{energyterms}, we present these energy terms for $^{208}\text{Pb}$ obtained in both 
the RMF and STF methods. In the RMF calculations, since the relativistic kinetic energy term 
cannot be directly obtained from the nucleon wave functions, we define it based on the
single-particle energy as follows:
\begin{eqnarray}
	E_{k}=\sum_{\alpha}^{\text{occ}} \left[w_{\alpha} \left(2j_{\alpha}+1\right)E_{\alpha}\right]
	-\int d^{3} r~\left[g_\omega \omega\left(n_p+n_n\right)
	+\frac{1}{2} g_\rho \rho\left(n_p-n_n\right)+ e A n_p \right].
\end{eqnarray}

\textcolor{black}{The density term associated with the $a$ meson is given by
	\begin{eqnarray}
		E_{a}=\int \frac{1}{2} g_a \braket{a} n_a ~d^{3}r,
	\end{eqnarray}
	where the corresponding source density $n_a$ can be identified in Equation~(\ref{STF energy}). }

In the RMF approach, the various energy terms of TM1m* and TM1m are very close to each other.
We note the differences in $E_{k}$ and $E_{\sigma}$ are related to the scalar field $\sigma$,
which may be attributed to different values of $m_{\sigma}$ used in the TM1m* and TM1m parameter sets.
On the other hand, the main difference between TM1m* and TM1e lies in the effective nucleon mass. 
Compared to the TM1e model, the TM1m* model has a larger effective mass $M^{\ast}$ and a weaker 
attractive potential $g_{\sigma}\sigma$. According to the analysis in the covariant density functional theory, the saturation 
mechanism relies on a delicate balance between the attractive $\sigma$ field and the 
repulsive $\omega$ field~\cite{SHARMA1994110}. Since the negative attractive potential in the TM1m* model
is relatively weak, the repulsive potential $g_{\omega}\omega$ is also reduced. These differences 
result in discrepancies in the relativistic kinetic energy $E_{k}$, as well as in the density terms 
$E_{\sigma}$ and $E_{\omega}$. The delicate balance among various energy contributions ensures that 
total energies in the TM1m and TM1e models are close to the experimentally required value.
Similar distinctions between TM1m* and TM1 occur, since the results in TM1e and TM1 are nearly identical.
The NL3 and TM1 models exhibit similar saturation properties and both provide good descriptions 
of finite nuclei. However, the most notable differences between these two models are found in the 
nonlinear energy terms associated with the $\sigma$ and $\omega$ mesons. This is primarily due to 
the differences in the coupling constants of nonlinear terms, as shown in Table~\ref{parameter}. 
The NL3 model uses a negative value for $g_3$, which has a smaller impact on the ground-state properties 
of finite nuclei. However, a negative $g_3$ may make the $\sigma$ meson equation asymptotically 
unstable and cause instabilities in calculations of the equation of state at very high 
densities~\cite{Waldhauser_1988, reinhard_1988}. Additionally, the NL3 model lacks a fourth-order 
term for the $\omega$ meson, causing its vector repulsive potential to increase too rapidly 
with increasing density. This leads to a stiffer equation of state in the NL3 model, 
which exceeds the constraints from heavy-ion collision (HIC) experiments~\cite{Kumar_2024}. 
The maximum mass of neutron stars obtained in the NL3 model, by solving the Tolman--Oppenheimer--Volkoff (TOV)
equation, is found to be as large as $2.77~M_{\odot}$~\cite{Miyatsu_2013, Kumar_2021}.
The results obtained in the SFHo model are comparable to those in the TM1m* model, as both models
exhibit similar saturation properties. The main differences lie in the attractive and repulsive 
potentials. The nonlinear term associated with the $\rho$ meson in the SFHo model has a small impact
and can therefore be neglected.

\begin{table}[H]
	\caption{Analysis of various energy terms for $^{208}\text{Pb}$ obtained using the RMF approach 
		and the STF approximation. $E_{k}$ represents the relativistic kinetic energy term of nucleons. 
		The density terms associated with the $\sigma$, $\omega$, $\rho$, and $A$ fields are denoted 
		by $E_{\sigma}$, $E_{\omega}$, $E_{\rho}$, and $E_{A}$, respectively. 
		The nonlinear terms of the $\sigma$ and $\omega$ mesons are respectively denoted by
		$E_{\sigma}^{\text{nl}}$ and $E_{\omega}^{\text{nl}}$, while $E_{\omega\rho}^{\text{cpl}}$ 
		represents the $\omega$-$\rho$ coupling term. 
		$E_{\text{total}}$ and $E/A$ are the total energy and energy per nucleon, respectively. 
		The energy terms are all given in $\text{MeV}$. The values in parentheses reflect 
		the differences between the STF and RMF results for each energy term. 
		The nonlinear term associated with the $\rho$ meson in the SFHo model is too small 
		and has therefore been omitted. Additionally, the center-of-mass correction and 
		pairing energy have been incorporated into the total energy.}
	\label{energyterms}
	\footnotesize
	\begin{adjustwidth}{-\extralength}{0cm}
		\centering
		\newcolumntype{C}{>{\centering\arraybackslash}X}
		\newcolumntype{L}{>{\raggedright\arraybackslash}X}
		\newcolumntype{R}{>{\raggedleft\arraybackslash}X}
		\begin{tabularx}{\fulllength}{CLRRRRRR}
			\toprule
			&         & \textbf{TM1m*}   & \textbf{TM1m}  & \textbf{SFHo}  & \textbf{TM1e}	& \textbf{TM1}   & \textbf{NL3}         \\
			\midrule
			\multirow{11}{*}{RMF} 
			& $E_{k}$                  & 166,769.60  & 166,089.04& 162,865.60&142,452.45  & 143,515.32 & 139,231.87  \\
			& $E_\sigma$               &  15,992.34  & 16,354.71 & 18,091.24 & 27,858.89  & 27,318.40  & 29,462.69    \\
			& $E_\omega$               & 11,116.80   & 11,403.13 & 13,079.89 & 22,986.98  & 22,508.88  & 24,682.22    \\
			& $E_\sigma^{\text{nl}}$   &  $-$1382.97  & $-$1439.95 & $-$1309.33 & $-$1325.01  & $-$1279.22  & $-$628.92    \\
			& $E_\omega^{\text{nl}}$   &   0.00     & 0.0005   & $-$4.80    &  500.71   & 475.07    &    0.00     \\
			& $E_\rho$                 &   123.10   &  124.82  & 108.20   & 116.78    & 109.43    & 103.85      \\
			& $E_A$                    &   820.30   &  822.60  & 840.33   &  819.93   &  823.42   & 827.17       \\
			& $E_{\omega\rho}^{\text{cpl}}$ & 39.32 &   40.88  & 29.53    & 60.89     &    0.00   &   0.00        \\
			& $E_{\text{total}}$       & 193,473.30  & 193,390.03& 193,695.47& 193,466.42 & 193,466.11 & 193,673.69  \\
			& $E/A$                    & $-$7.84      & $-$8.24    & $-$7.83    &  $-$7.87    &   $-$7.87   & $-$7.88          \\
			\midrule
			\multirow{19}{*}{STF} & $E_{k}$    & \begin{tabular}[c]{@{}r@{}}165,995.23\\ ($-$774.37)\end{tabular}                     &  \begin{tabular}[c]{@{}r@{}}165,157.89\\ ($-$931.15)\end{tabular}                    & \begin{tabular}[c]{@{}r@{}}162,479.36\\ ($-$386.24)\end{tabular}		&\begin{tabular}[c]{@{}r@{}}143,555.73\\ (1103.28)\end{tabular} &        \begin{tabular}[c]{@{}r@{}}144,583.83\\ (1068.51)\end{tabular}              & \begin{tabular}[c]{@{}r@{}}141,928.88\\ (2697.02)\end{tabular}    \\
			& $E_\sigma$ &        \begin{tabular}[c]{@{}r@{}}16,371.93\\ (379.59)\end{tabular}               &  \begin{tabular}[c]{@{}r@{}}16,818.50\\ (463.79)\end{tabular}                    & \begin{tabular}[c]{@{}r@{}}18,258.43\\ (167.19)\end{tabular}	&\begin{tabular}[c]{@{}r@{}}27,250.99\\ ($-$607.89)\end{tabular}   		&    \begin{tabular}[c]{@{}r@{}}26,730.02\\ ($-$588.38)\end{tabular}                   & \begin{tabular}[c]{@{}r@{}}28,076.58\\ ($-$1386.11)\end{tabular}     \\
			& $E_\omega$ &         \begin{tabular}[c]{@{}r@{}}11,405.32\\ (288.52)\end{tabular}              &        \begin{tabular}[c]{@{}r@{}}11,752.04\\ (348.91)\end{tabular}               & \begin{tabular}[c]{@{}r@{}}13,208.84\\ (128.95)\end{tabular}		& \begin{tabular}[c]{@{}r@{}}22,474.92\\ ($-$512.06)\end{tabular}		&    \begin{tabular}[c]{@{}r@{}}22,014.45\\ ($-$494.43)\end{tabular}                   & \begin{tabular}[c]{@{}r@{}}23,445.06\\ ($-$1237.16)\end{tabular}         \\
			& $E_\sigma^{\text{nl}}$ &          \begin{tabular}[c]{@{}r@{}}$-$1386.61\\ ($-$3.64)\end{tabular}             &    \begin{tabular}[c]{@{}r@{}}$-$1449.11\\ ($-$9.16)\end{tabular}                   & \begin{tabular}[c]{@{}r@{}}$-$1300.66\\ (8.67)\end{tabular}			& \begin{tabular}[c]{@{}r@{}}$-$1237.30\\ (87.71)\end{tabular}&  \begin{tabular}[c]{@{}r@{}}$-$1194.37\\ (84.85)\end{tabular}                     & \begin{tabular}[c]{@{}r@{}}$-$663.57\\ ($-$34.65)\end{tabular}     \\
			& $E_\omega^{\text{nl}}$ & \begin{tabular}[c]{@{}r@{}} 0.00 \\ (0.00)\end{tabular}                        &    \begin{tabular}[c]{@{}r@{}}0.00\\ (0.00)\end{tabular}                    & \begin{tabular}[c]{@{}r@{}}$-$4.49\\ (0.31)\end{tabular}	 & \begin{tabular}[c]{@{}r@{}}437.48\\ ($-$63.23)\end{tabular}&          \begin{tabular}[c]{@{}r@{}}414.95\\ ($-$60.12)\end{tabular}                 & \begin{tabular}[c]{@{}r@{}}0.00\\ (0.00)\end{tabular}     \\
			& $E_\rho$ &          \begin{tabular}[c]{@{}r@{}}135.23\\ (12.13)\end{tabular}               &        \begin{tabular}[c]{@{}r@{}}139.01\\ (14.19)\end{tabular}                 & \begin{tabular}[c]{@{}r@{}}117.64\\ (9.44)\end{tabular}& \begin{tabular}[c]{@{}r@{}}124.98\\ (8.20)\end{tabular}&  \begin{tabular}[c]{@{}r@{}}113.85\\ (4.42)\end{tabular}                       & \begin{tabular}[c]{@{}r@{}}105.69\\ (1.84)\end{tabular}     \\
			& $E_A$ &       \begin{tabular}[c]{@{}r@{}}816.16\\ ($-$4.15)\end{tabular}                  & 	 \begin{tabular}[c]{@{}r@{}}816.24\\ ($-$6.37)\end{tabular}                       & \begin{tabular}[c]{@{}r@{}}833.75\\ ($-$6.58)\end{tabular}& \begin{tabular}[c]{@{}r@{}}807.26\\ ($-$12.67)\end{tabular} & \begin{tabular}[c]{@{}r@{}}810.17\\ ($-$13.25)\end{tabular}                        & \begin{tabular}[c]{@{}r@{}}808.37\\ ($-$18.80)\end{tabular}       \\
			& $E_{\omega\rho}^{\text{cpl}}$     &   \begin{tabular}[c]{@{}r@{}}43.59\\ (4.28)\end{tabular}                         &       \begin{tabular}[c]{@{}r@{}}46.21\\ (5.33)\end{tabular}                   & \begin{tabular}[c]{@{}r@{}}31.28\\ (1.75)\end{tabular}	& \begin{tabular}[c]{@{}r@{}}64.05\\ (3.16)\end{tabular}							& \begin{tabular}[c]{@{}r@{}}0.00\\ (0.00)\end{tabular}                        & \begin{tabular}[c]{@{}r@{}}0.00\\ (0.00)\end{tabular}    \\
			&  $E_{\text{total}}$      &   \begin{tabular}[c]{@{}r@{}}193,380.85\\ ($-$97.64)\end{tabular}                   &   \begin{tabular}[c]{@{}r@{}}193,280.77\\ ($-$114.45)\end{tabular}                   & \begin{tabular}[c]{@{}r@{}}193,624.16\\ ($-$76.50)\end{tabular}		& \begin{tabular}[c]{@{}r@{}}193,478.11\\ (6.50)\end{tabular}& \begin{tabular}[c]{@{}r@{}}193,472.89\\ (1.59)\end{tabular}                     & \begin{tabular}[c]{@{}r@{}}193,701.01\\ (22.14)\end{tabular}     \\
			&  $E/A$       &     $-$8.28       &   $-$8.77      & $-$8.17  &    $-$7.82            &         $-$7.84                & $-$7.75      \\
			\bottomrule
		\end{tabularx}
	\end{adjustwidth}
\end{table}

In the STF approximation, the energy terms in the six models exhibit a similar pattern as 
observed in the RMF approach. Our primary focus lies on the values in parentheses, which represent 
the differences between the STF and RMF results for each energy term. These differences reveal 
what causes the same model to yield divergent outcomes under the two frameworks.
The values in parentheses show significant discrepancies in the relativistic kinetic energy $E_{k}$,
as well as in the density terms $E_{\sigma}$ and $E_{\omega}$ associated with $\sigma$ and $\omega$ mesons, 
all of which are related to the Dirac equation. This may be due to different methods used to handle the 
Dirac equation and extract the kinetic energy within the two approaches.
In the RMF approach, the single-particle energy of nucleons is self-consistently obtained by 
solving the Dirac equation. In contrast, under the STF approximation, nucleons are treated as moving 
in locally constant fields, and the kinetic energy is calculated in a manner analogous to that 
in homogeneous nuclear matter. It is likely these models can be divided into two groups.
In the TM1m*, TM1m, and SFHo models, the values in parentheses exhibit similar features
with the same sign, particularly for $E_{k}$, $E_{\sigma}$, and $E_{\omega}$. 
In contrast, in the TM1e, TM1, and NL3 models, the sign of these values exhibits an opposite trend.

\subsection{Charge Radius}
\label{subsection:RC}

\begin{figure}[H]
	\begin{adjustwidth}{-\extralength}{0cm}
		\centering
		\includegraphics[width=1.0\linewidth]{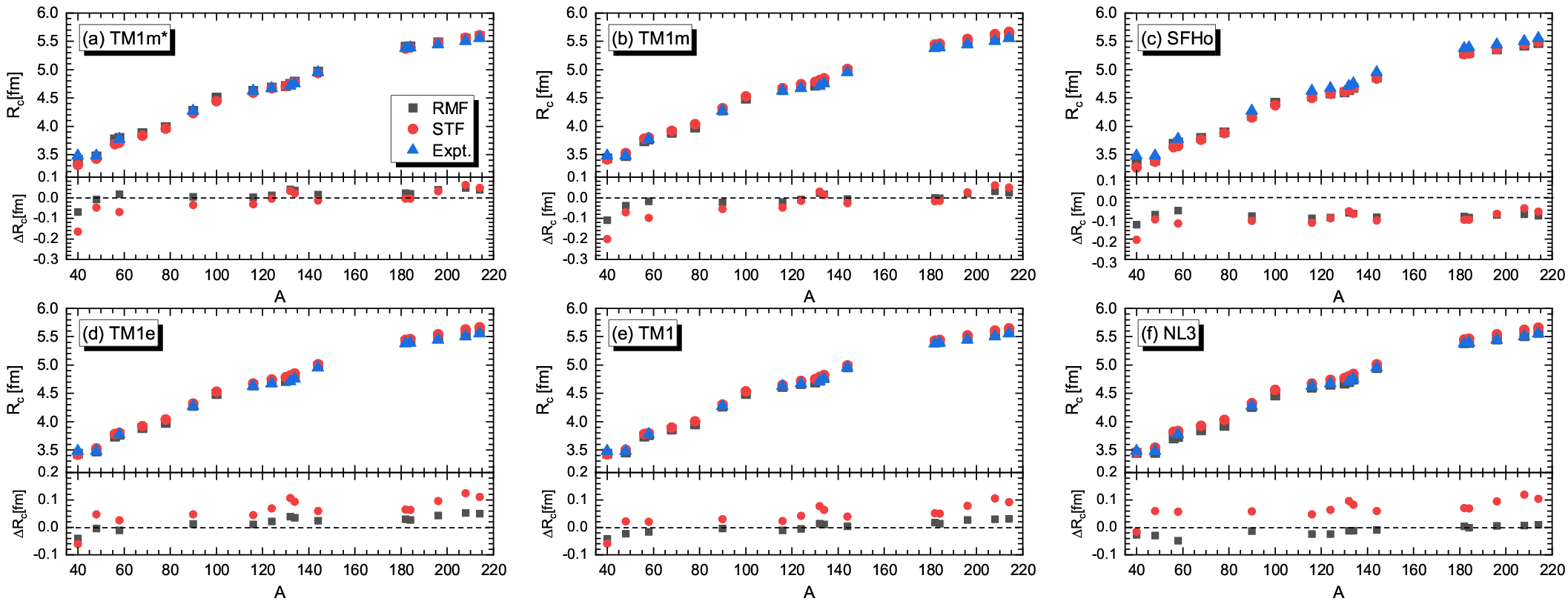}
	\end{adjustwidth}
	\caption{The same as Figure~\ref{31BE} but for the charge radius $R_{c}$. The experimental data are
		taken from Ref.~\cite{ANGELI201369}. $\Delta R_{c}$ denotes the difference between the 
		model prediction and experimental data, defined by $\Delta R_{c}=R_{c, \text{RMF/STF}}-R_{c, \text{Expt.}}$.}
	\label{32RC}
\end{figure}

To compare the charge radii between the RMF and STF approaches, we plot in Figure~\ref{32RC} the charge 
radii for the 19 selected nuclei as a function of the mass number $A$. The experimental data are taken 
from Ref.~\cite{ANGELI201369}, although some values are absent.

Comparing Figure~\ref{32RC}a,b, we find that, in the TM1m* and TM1m models, the charge radii obtained using the RMF and STF approaches are nearly identical and very close to the experimental values. 
In the case of the SFHo model, as shown in Figure~\ref{32RC}c, the theoretical predictions from both the RMF and STF approaches are slightly lower than the experimental values. 
On the other hand, in the TM1e, TM1, and NL3 models shown in Figure~\ref{32RC}d--f, the charge radii obtained using the STF approximation are obviously larger than those 
in the RMF approach. This difference may be attributed to the different density distributions of the 
two approaches. 
The results in the TM1e model are slightly larger than those in the TM1 model. 
This might be due to the fact that TM1e has a smaller slope parameter $L=40\,\text{MeV}$, 
which results in a larger proton radius. 

\textcolor{black}{We have calculated the charge radii using Equation~(\ref{RC}) proposed by Sugahara and 
	Toki~\cite{SUGAHARA1994557}. Alternatively, Typel and Wolter have suggested another formula, 
	Equation~(45) in Ref.~\cite{typel1999}, for calculating charge radii. 
	We have compared the results obtained with these two formulas. The comparison shows that their 
	mutual discrepancies are much smaller than those between theoretical predictions and 
	experimental data. The qualitative conclusion is therefore unaffected. }

Recently, more attention has been paid to the relationship between the symmetry 
energy slope $L$ and the neutron-skin thickness $\Delta R_{np}$~\cite{Zhang_2014,Shen_2020}. 
Generally, a larger value of $L$ corresponds to a higher pressure of neutron matter and a greater
neutron-skin thickness. Accurate measurements of neutron-skin thickness deduced from parity-violating
electron scattering suggests a small neutron-skin thickness of $^{48}\text{Ca}$ by CREX~\cite{adhi22}
and a large neutron-skin thickness of $^{208}\text{Pb}$ by PREX-2~\cite{adhi21}.
In Table~\ref{drnp}, we present the resulting neutron-skin thickness, $\Delta R_{np}=R_n-R_p$, 
for $^{48}\text{Ca}$ and $^{208}\text{Pb}$ obtained using the RMF approach and the STF approximation,
and compared with experimental data from CREX~\cite{adhi22} and PREX-2~\cite{adhi21}.
It is noteworthy that the TM1 and NL3 models, characterized by large values of symmetry energy slope $L$,
predict systematically greater neutron-skin thicknesses than the other models with small $L$.
Furthermore, the results obtained using the STF approximation are smaller than those using the RMF approach.
This difference is likely attributed to the different density distributions of the two approaches.
Compared to the experimental data, we find that none of the models can simultaneously predict the 
neutron-skin thicknesses for $^{48}\text{Ca}$ and $^{208}\text{Pb}$ within the experimental error range.
Currently, there are still large uncertainties in the neutron-skin thicknesses obtained 
from experimental measurements.

\begin{table}[H]
	\caption{Neutron-skin thickness (in fm) obtained using the RMF approach and the STF approximation. 
		The experimental data for $^{48}\text{Ca}$ and $^{208}\text{Pb}$ are taken from Ref.~\cite{adhi22}
		and~Ref.~\cite{adhi21}, respectively.}
	\label{drnp}
	\setlength{\cellWidtha}{\textwidth/4-2\tabcolsep-0.3in}
	\setlength{\cellWidthb}{\textwidth/4-2\tabcolsep-0.8in}
	\setlength{\cellWidthc}{\textwidth/4-2\tabcolsep+0.8in}
	\setlength{\cellWidthd}{\textwidth/4-2\tabcolsep+0.3in}
	\scalebox{1}[1]{\begin{tabularx}{\textwidth}{>{\centering\arraybackslash}m{\cellWidtha}>{\raggedright\arraybackslash}m{\cellWidthb}>{\centering\arraybackslash}m{\cellWidthc}>{\centering\arraybackslash}m{\cellWidthd}}
			\toprule
			\multicolumn{1}{l}{} &       & \boldmath{\textbf{$^{48}\text{Ca}$}}                   & \boldmath{\textbf{$^{208}\text{Pb}$}}         \\
			\midrule
			\multirow{6}{*}{RMF} & TM1m* & 0.1934                                                  & 0.1782        \\
			& TM1m  & 0.1807                                                  & 0.1659        \\
			& SFHo  & 0.1992                                                  & 0.1931        \\
			& TM1e  & 0.1712                                                  & 0.1579        \\
			& TM1   & 0.2274                                                  & 0.2709        \\
			& NL3   & 0.2271                                                  & 0.2808        \\
			\midrule
			\multirow{6}{*}{STF} & TM1m* & 0.117                                                   & 0.104         \\
			& TM1m  & 0.098                                                   & 0.082         \\
			& SFHo  & 0.135                                                   & 0.129         \\
			& TM1e  & 0.121                                                   & 0.103         \\
			& TM1   & 0.184                                                   & 0.211         \\
			& NL3   & 0.189                                                   & 0.222         \\
			\midrule
			\multicolumn{2}{c}{Experimental data}    & \multicolumn{1}{l}{$0.121 \pm 0.026$ (exp) $\pm$ 0.024 (model)} & \mbox{$0.283 \pm 0.071 $} \\
			&  &      \mbox{($0.071 \sim 0.171$)}             & \mbox{($0.212 \sim 0.354$)}  \\
			\bottomrule
	\end{tabularx}}
\end{table}

\subsection{Nucleon Density Distribution}
\label{subsection:DD}

The density distributions of protons and neutrons provide crucial insights into the internal structure 
of finite nuclei. We now turn our attention to the discussion of density distributions for 
two typical nuclei, $^{40}\text{Ca}$ and $^{208}\text{Pb}$.

\begin{figure}[H]
	\begin{adjustwidth}{-\extralength}{0cm}
		\centering
		\includegraphics[width=1.0\linewidth]{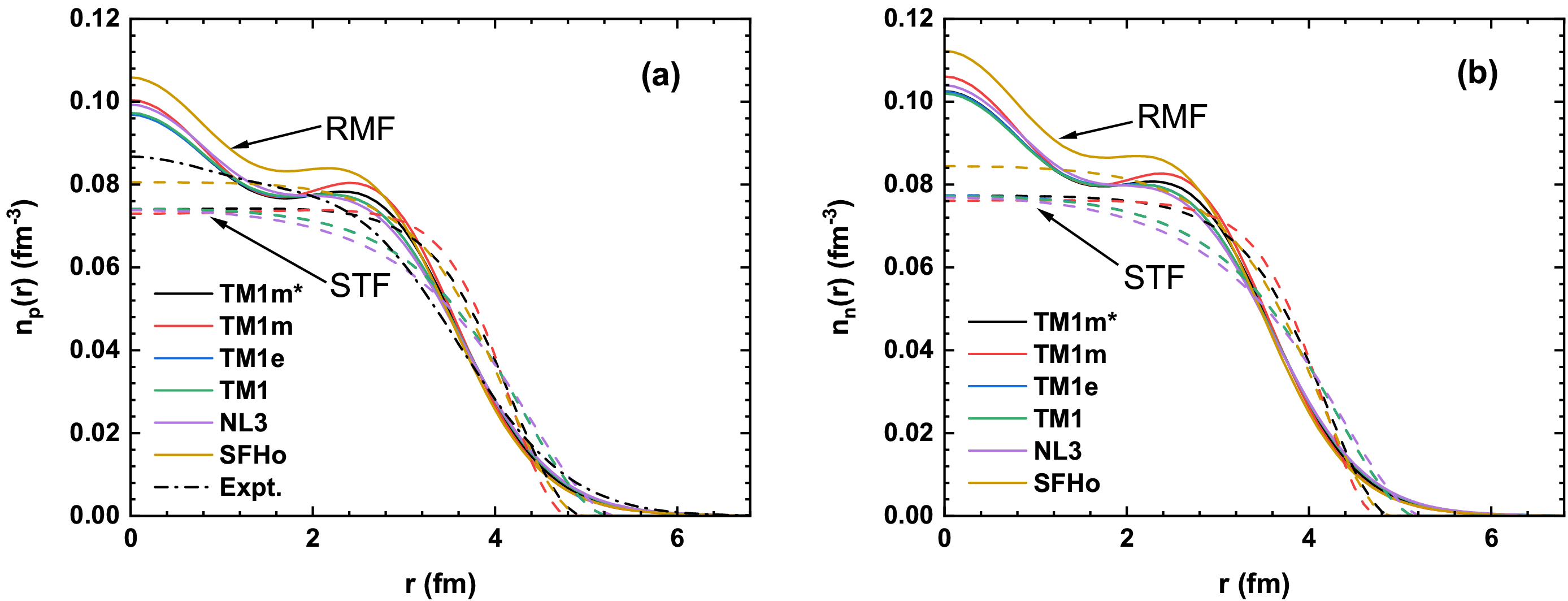}
	\end{adjustwidth}
	\caption{Density distributions of (\textbf{a}) protons and (\textbf{b}) neutrons for $^{40}\text{Ca}$. 
		\textcolor{black}{The experimental data are the charge density distribution obtained from elastic electron 
			scattering~\cite{DEVRIES1987495}.}}
	\label{33ca}
\end{figure}

In Figure~\ref{33ca}, we display the density distributions of protons and neutrons for $^{40}\text{Ca}$. 
The curves obtained using the RMF approach exhibit pronounced fluctuations and spatially diffuse distributions, 
with minimal differences observed in the tail regions across different parameter sets. 
The density distributions of the TM1e and TM1 models show nearly identical, while the TM1m*
curves differ with them only in the middle region (\mbox{$2\,\text{fm} <r<3\,\text{fm}$}). 
Similarly, the curves of the TM1m and NL3 models exhibit slight deviations from that of the TM1e model.
However, the SFHo model yields enhanced central density compared to other models, which may be related
to its larger saturation density, as shown in Table~\ref{properties}. 
This characteristic also contributes to the systematically smaller charge radii predicted by the SFHo model.

According to Equation~(\ref{density_RMF}), the complex structure of the density distribution obtained from 
the RMF approach originates from the superposition of squared wave function for all occupied states. 
Different parameter sets yield distinct wave functions $\phi_{\alpha}$ for the same single-particle state. 
In open-shell nuclei, the occupation probabilities $w_{\alpha}$ of occupied states also vary across 
parameter sets. This inconsistency in occupation probabilities further enhances the parameter dependence 
of the density distributions. 
\textcolor{black}{Furthermore, the density distributions calculated from the superposition of squared wave 
	function for all occupied states retain the oscillatory features inherent in the individual radial 
	wave functions, giving rise to pronounced oscillations in the RMF density distributions. 
	These oscillations, however, substantially exceed the fluctuation amplitudes observed in the 
	charge density distributions deduced from elastic electron scattering experiments~\cite{DEVRIES1987495}.}

Within the STF approximation, the density distributions exhibit relatively smooth variations and have 
three distinctive characteristics compared to the RMF results: 
(1) lower densities in the central region; 
(2) slower attenuation with increasing radius, eventually exceeding the RMF densities and decreasing rapidly;
and (3) reduced spatial diffusion in the tail region, where greater discrepancies emerge between 
different parameter sets.
\textcolor{black}{The STF approximation produces relatively smooth distributions because it treats 
	nucleons as a locally uniform Fermi gas subject to slowly varying mean fields.
	The fundamental difference between the STF approximation and the RMF approach is in how nucleons 
	are handled, which leads to the pronounced disparities between the sets of density distributions.
	In terms of predictive capability, although the oscillations in the RMF densities deviate from 
	the experimental charge density profiles, the RMF approach successfully reproduces various 
	properties and deformation characteristics of finite nuclei. In contrast, the STF 
	approximation yields reliable ground-state bulk properties but cannot predict microscopic 
	quantities such as single-particle spectra.}

\begin{figure}[H]
	\begin{adjustwidth}{-\extralength}{0cm}
		\centering
		\includegraphics[width=0.95\linewidth]{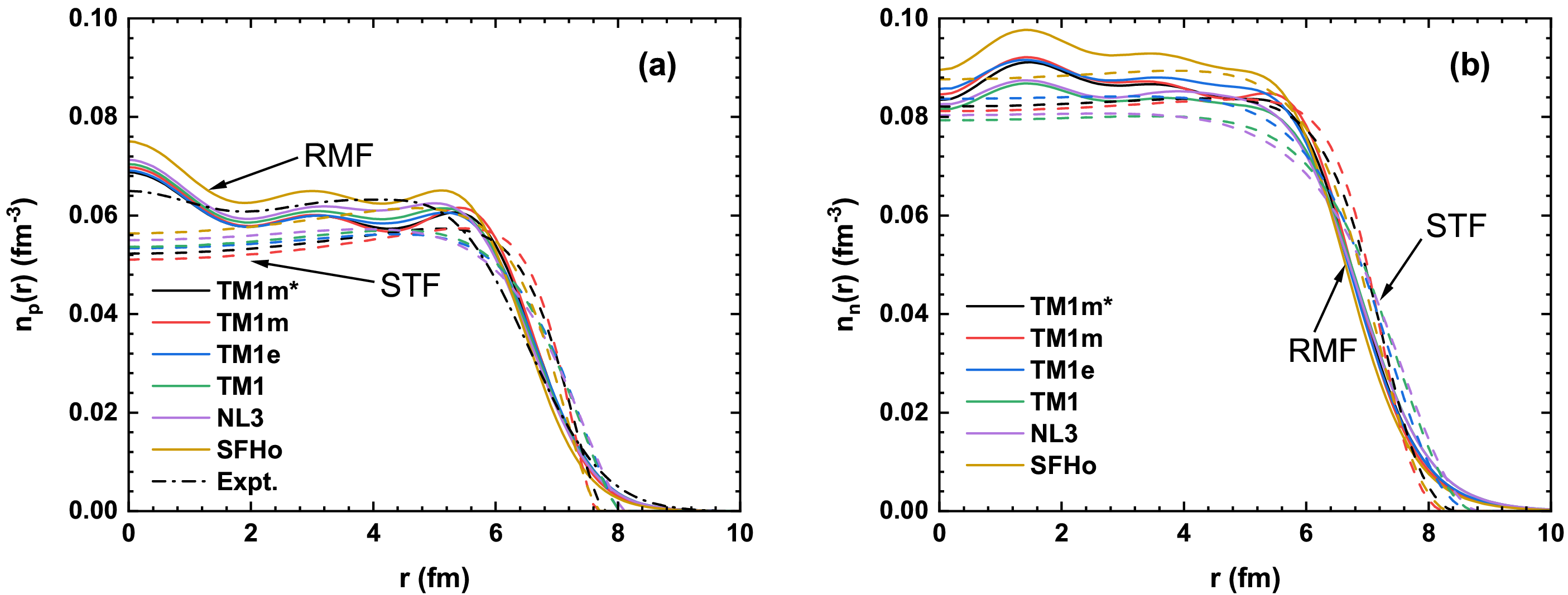}
	\end{adjustwidth}
	\caption{Density distributions of (\textbf{a}) protons and (\textbf{b}) neutrons for $^{208}\text{Pb}$. 
		\textcolor{black}{The experimental data are the charge density distribution obtained from elastic 
			electron scattering~\cite{DEVRIES1987495}.}}
	\label{34pb}
\end{figure}

In Figure~\ref{34pb}, the density distributions of protons and neutrons for $^{208}\text{Pb}$ are presented.
It is shown that, for $^{208}\text{Pb}$, the curves obtained using the RMF and STF approaches exhibit 
similar features to those of $^{40}\text{Ca}$. However, more pronounced differences among these models 
are observed in the curves of $^{208}\text{Pb}$. The curves of the TM1e and TM1 models no longer overlap, 
particularly in the neutron distribution. The curves of the neutron distribution 
show much higher sensitivity to the symmetry energy slope $L$ than that of the proton distribution. 
This fact supports the theoretical prediction that the value of $L$ predominantly affects the neutron radius.
\textcolor{black}{For $^{208}\text{Pb}$, the density distributions exhibit more pronounced oscillations
	compared to $^{40}\text{Ca}$.} 
Additionally, in Figure~\ref{34pb}a, the proton density distribution within the STF approximation 
exhibits an inward indentation in the central region of the nucleus, which is mainly attributed to 
the Coulomb repulsion~\cite{BOGUTA197722}. This effect occurs in heavy nuclei because their larger Coulomb 
repulsion causes a redistribution of protons, leading to a lower density 
in the central region compared to the surrounding areas.

\section{Conclusions}
\label{Conclusions}

In this work, we have made a detailed comparison between the STF approximation and the RMF approach
for describing finite nuclei. We have employed the \textcolor{black}{covariant density functional theory}
with nonlinear terms for $\sigma$ and $\omega$ mesons, as well as an $\omega$-$\rho$ coupling term. 
Besides considering several typical RMF parameter sets, we have provided a new TM1m* parameterization. 
The TM1m* features a high effective nucleon mass ratio, $M^{\ast}/M \sim 0.8$, identical to that 
of the TM1m model, but with better descriptions for finite nuclei. 
Recent studies have indicated that a high effective mass is helpful for the shock evolution and 
explosion processes in core-collapse supernovae. The other saturation properties of TM1m*  
are very similar to those of the TM1m and TM1e models. On the other hand, the TM1e and TM1 models
have lower effective nucleon mass ratio, $M^{\ast}/M \sim 0.63$. The difference between TM1e and TM1 
is reflected in the symmetry energy and its slope. It is well known that the slope parameter
of the symmetry energy can significantly affect both the neutron-star radius and the neutron-skin 
thickness of finite nuclei. In addition, two popular models, NL3 and SFHo, have also been employed
in our calculations for comparison. The NL3 model has been shown to perform well in describing nuclear properties. The SFHo model has been widely used in astrophysical simulations and has a large 
effective mass ratio, $M^{\ast}/M \sim 0.76 $, similar to the TM1m* models.

To systematically analyze the differences between the STF approximation and the RMF approach for 
the descriptions of finite nuclei, we have calculated the ground-state properties of 19 doubly magic 
and semi-magic nuclei that span a wide range of the nuclide chart. The experimental data of 
the corresponding nuclei have been used to examine these results. Within the RMF approach, the models
used could provide reasonable good results for finite nuclei, whereas the TM1m model exhibits relatively
large discrepancies between theoretical predictions and experimental values.
This is because the ground-state properties of finite nuclei have not been included in the fitting 
process for the TM1m parameter set, whereas the TM1m* parameterization considered and improved its
description of finite nuclei. On the other hand, within the STF approximation, models with larger
effective mass, such as the TM1m*, TM1m, and SFHo models, generally predict results lower than 
the experimental values. In contrast, the TM1e and TM1 models show better agreement with 
the experimental data, while the NL3 model exhibits slightly higher predictions.  
Notably, when using the same parameter set, the variations in predictions between the two methods 
primarily arise from different treatments for the relativistic kinetic term and the density terms
related to the Dirac equation. 

For the charge radius of finite nuclei, most parameter sets can provide reasonably good predictions 
in the RMF approach. It has been found that the TM1e model predicts slightly larger charge radii 
than the TM1 model, which might be attributed to the smaller slope parameter of the TM1e model.
In the STF approximation, the TM1m* and TM1m parameter sets can reproduce the experimental charge 
radii, whereas the results from the SFHo model remain lower than the experimental values.
Furthermore, in the TM1e, TM1, and NL3 models, the charge radii obtained using the STF approximation
are obviously larger than those in the RMF approach. 

We have also discussed the nucleon density distributions for two typical nuclei, $^{40}\text{Ca}$ 
and $^{208}\text{Pb}$. The density distributions obtained using the RMF approach exhibit
pronounced fluctuations and spatially diffuse characteristics, with minor differences 
observed in the tail regions across different parameter sets.
In contrast, the density distributions obtained through the STF approximation exhibit 
relatively smooth variations, showing lower values in the central regions and weaker tail diffuse. 
These features are particularly pronounced in the results of $^{208}\text{Pb}$. 
Furthermore, the discrepancy between the TM1e and TM1 curves demonstrates that 
neutron distributions are more sensitive to the symmetry energy slope $L$ than 
the proton distributions.
\vspace{6pt} 


\authorcontributions{Conceptualization, H.S.; methodology, S.L. and J.H.; software, S.L.; validation, S.L., H.S., and J.H.; investigation, S.L., H.S., and J.H.; supervision, H.S. All authors have read and agreed to the published version of the manuscript.}

\funding{ This research was partially supported by the National Natural Science Foundation of China under Grants Nos. 12175109 and 12475149.}

\dataavailability{The original contributions presented in this study are included in the article. Further inquiries can be directed to the corresponding authors.} 


\conflictsofinterest{The authors declare no conflicts of interest.} 
	\begin{adjustwidth}{-\extralength}{0cm}
	
	\reftitle{References}

	\PublishersNote{}
	\end{adjustwidth}
\end{document}